\documentclass[letterpaper, 10 pt, conference]{ieeeconf}  
\usepackage{graphicx}   

 \usepackage{url}
\IEEEoverridecommandlockouts                              
\title{\LARGE \bf
Objective Prediction of Tomorrow’s Affect Using Multi-Modal Physiological Data and Personal Chronicles: A Study of Monitoring College Student Well-being in 2020
}

\author{Salar Jafarlou $^{1}$, Jocelyn Lai $^{2}$,  Zahra Mousavi $^{2}$, Sina Labbaf $^{1}$,\\ Ramesh C. Jain $^{1}$, Nikil D. Dutt $^{1}$, Jessica L. Borelli $^{3}$,  Amir M. Rahmani $^{1,4}$

    \thanks{$^{1}$University of California Irvine, Department of Computer Science }%
    \thanks{$^{2}$University of California, Irvine, Department of Psychological Science }%
    \thanks{$^{3}$University of California Irvine, Department of Cognitive Sciences }%
    \thanks{$^{4}$University of California Irvine, School of Nursing }%

    }

\begin{document}

\maketitle
\thispagestyle{empty}
\pagestyle{empty}

\begin{abstract}

Monitoring and understanding affective states are important aspects of healthy functioning and treatment of mood-based disorders. Recent advancements of ubiquitous wearable technologies have increased the reliability of such tools in detecting and accurately estimating mental states (e.g., mood, stress, etc.), offering comprehensive and continuous monitoring of individuals over time. Previous attempts to model an individual’s mental state were limited to subjective approaches or the inclusion of only a few modalities (i.e., phone, watch). Thus, the goal of our study was to investigate the capacity to more accurately predict affect through a fully automatic and objective approach using multiple commercial devices. Longitudinal physiological data and daily assessments of emotions were collected from a sample of college students using smart wearables and phones for over a year. Results showed that our model was able to predict next-day affect with accuracy comparable to state of the art methods. 

\end{abstract}

\section{INTRODUCTION}
National estimates of mental health suggest that as of 2019, 1 in 5 adults in the U.S. experience mental illness, with young adults at greater risk than their older counterparts \cite{cite1}. An important facet of mental health is affect, subjective and physiological experiences of positive and negative emotions \cite{cite2}. Affective disturbances and dysregulation - more specifically, overly heightened or dampened affect, prolonged negative affect, or instability in experienced affect are core facets of many types of psychopathology. Since these marks often precede the onset of psychopathology, they are considered to be risk factors for psychopathology \cite{cite3}. Monitoring and increased understanding of one’s affect contributes to the regulation of affect \cite{cite4}, and as such is a key component of many forms of intervention or approaches to manage affective disturbances \cite{cite5,cite6}. In monitoring one’s affect, patients and their provider can determine contextual factors relevant to experienced emotions and moods, and in turn develop a plan through medication or therapy to facilitate cognition and behavior in such a way that will reduce disturbances. Traditional approaches rely on the patient self-reporting or monitoring through diary or logging and require assistance from a therapist or provider. 
The prevalence of technology, advances in Internet-of-things (IoT), Wearable IoT (WIoT), and machine learning offer an easily accessible and less burdensome approach to objectively monitor and predict affect, which can result in a better understanding of an individual’s affect \cite{cite8,cite14}. These advancements have opened new gateways in monitoring an individual’s different physical and mental health aspects in a continuous and uninterrupted way. While modeling behavioral mental health markers through smartphones already existed \cite{cite9}, the research community gradually started to utilize wearables technologies as additional modalities to assess physiology (e.g., autonomic nervous systems activity in response to mental health stressors) and develop more in-depth analyses of behavior (e.g., sleep, physical activity, etc.). This resulted in more holistic information about an individual's daily life \cite{cite10}.
One notable study using a holistic approach is (SNAPSHOT) \cite{cite13} that assessed participant daily life for one month using two wristbands and a smartphone per user. Authors found that some indicators of an individual’s well-being (e.g., mood, stress and health) could be modeled using a set of objective behavioral and biological features. In addition to objective assessments, participants in this study completed self-reported assessments about their activities and interactions. Subjective data assessments require a user’s dedicated attention, which may result in increased burden on users and unsatisfactory experience, and thus be subject to higher rates of missing or inconsistent data. Furthermore, a majority of daily monitoring studies require participants to track or complete assessments over the course of a few weeks to a month \cite{cite7} which depending on when the study is conducted, may be inadequate in capturing long-term changes or responses to certain stressors and events. Taylor and colleagues \cite{cite6} achieved up to \% of accuracy in predicting next day mood using general models on SNAPSHOT dataset. 
However, there may still be varying events over a year or large-scale collective traumas that may relate to an individual’s experiences of emotion and mood \cite{cite20}. For example, the COVID-19 pandemic lockdown within the United States was initiated in March of 2020 and lasted through the first half of 2021, with variation in mandates across states. The duration of the pandemic has had large impacts on livelihood and social norms \cite{cite8}. During this watershed time, emerging adults, who often are at the height of forming their identities and social networks, may have experienced greater distress and adjustment in light of stay-at-home orders that mandated they isolate themselves in the home environment. Following individuals over the course of a year may offer a more comprehensive understanding of the fluctuations of affect in relation to contextual factors in daily life \cite{cite9}. 
Thus, the goal of this study is to address the aforementioned limitations of previous studies (e.g., time span and partial reliance on subjective assessment), by both expanding the time length of data collection to about 12-months during the eventful year of 2020 and also by limiting to exclusively use objective measurements to model users’ next day affect. Using only objective measures by commercial devices to predict an individual's affect can enable us to monitor mental health in a more continuous and convenient way; thus making mental health monitoring service accessible to more people. We monitored the life of 20 college students pre- and post COVID-19 lock-down. Our results demonstrate that it is feasible to predict an individual’s affect in a longer timeframe by only relying on objectively collected data while achieving up to 23\% of relative improvement in accuracy compared to the state-of-the-art.

\section{Affect Prediction Study}

In this section we discuss different aspects of collected data in this study, i.e., collection setup, some insight about features and labels, and the mood prediction modeling scheme. 

\subsection{Data Collection Setup}
To develop our fully objective models, we collected data from a sample of college students (N=20, 65\% female, Mage=19.80, SDage=1.0) as part of a larger study aimed at assessing personalized approaches to understanding mental health and well-being among emerging adults \cite{cite10}. Full-time college students between the ages of 18-22 who were unmarried, English fluent, and used an Android device as their primary phone were recruited from a large west-coast university to participate in this extensive longitudinal study conducted over a  12-month period. Data were collected throughout 2020, specifically during the COVID-19 pandemic. This study was approved by the Institutional Review Board prior to its inception. Participants provided written consent prior to their participation. 
College students were monitored over a 12-month period where they wore non-invasive commercial smart devices that captured physiology, sleep, and physical activity. Students also  downloaded a smartphone application that automatically detected activities and lifelogs (e.g., staying still, walking, in-vehicle, etc.), movement, and location. During this monitoring period, participants completed daily surveys using a separate smartphone application that prompted them to report their affect based on a list of 20 discrete emotion words at the end of the day. Participants’ daily self-reported affect was used to evaluate our predictive models.

\begin{figure}[!th]
    \centering
    \includegraphics[width=\linewidth]{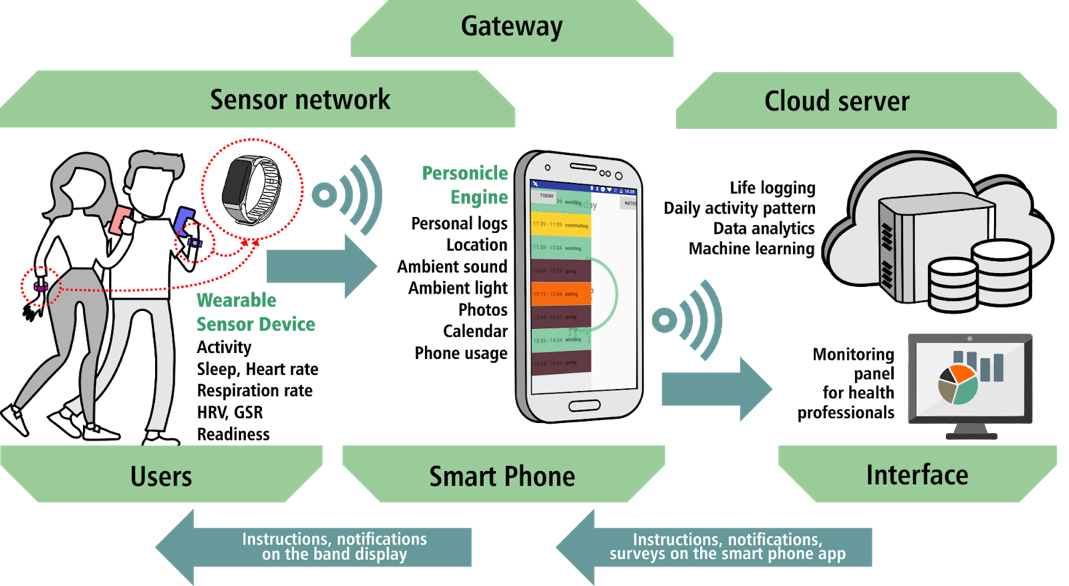}%
    \caption{Data collection setup; wearable devices collect targeted information through applications we developed and send the data to our cloud server by proxying the smartphone internet connection. Smartphones also are used to collect participant’s daily affect assessments.}
    \label{fig:setup}
\end{figure}

We utilized a number of services offered by ZotCare \cite{cite15} to collect different types of data. ZotCare is a dynamic and flexible multi-layer (sensor-smartphone-cloud) platform built in the University of California, Irvine (UCI) Institute for Future Health (IFH) which provides a variety of services necessary to run a human study trial in one place (Figure \ref{fig:setup} ). It is designed to support the entire life-cycle of a remote monitoring-intervention study from planning to deployment and analysis. It offers i) services to interface with wearable and portable sensors (both native and third party), ii) a multi-platform mobile app for two-way communication with users, collecting EMAs and self-reports, sending recommendations and reminders, and implementing remote interventions, iii) back-end services for storage, computation, control, and visualization, and iv) a web-based dashboard for workers, researchers, and end-users interact with the system. In this study we utilized ZotCare to collect data from participants. We installed custom services on the participants’ smartwatches to collect, store, and transfer data to the ZotCare server. Participants’ Oura ring data was also accessible through ZotCare using a server-to-server connection to Oura’s server. Subjective mood assessment was collected using ZotCare’s mobile app customized for this study.
\begin{table*}
\fontsize{7}{9}\selectfont
\renewcommand{\arraystretch}{1}
\begin{tabular}{ |p{0.6cm}|p{0.6cm}|p{1.5cm}|p{1.7cm}|p{1.7cm}|p{1cm}|p{1cm}|p{1.5cm}|p{0.6cm}|p{0.6cm}|p{1.5cm}|  }\hline

 \multicolumn{8}{|c|}{Ring} &   \multicolumn{2}{|c|}{Watch} &   \multicolumn{1}{|c|}{Phone} \\ \hline
 
 \multicolumn{2}{|c|}{Sleep (mins)} &  \multicolumn{2}{|c|}{Activity} &  \multicolumn{2}{|c|}{Metabolic} & \multicolumn{1}{|c|}{Calorie} & \multicolumn{1}{|c|}{Heart}& \multicolumn{1}{|c|}{Distance}& \multicolumn{1}{|c|}{Environment}& \multicolumn{1}{|c|}{Activity}\\ \hline

  Awake & REM & Stay Active & Meet Daily Activity Target & Avg MET(*) & MET Inactive & Calorie Active & Heart Rate & Distance & Pressure & Main Activity \\ \hline
  
  Light & Deep & Move Every Hour & Training Frequency & Minutes Low Activity & MET Low & Calorie Total & Heart Rate std & Runsteps & Pressure min & Key Activity \\ \hline

  Total &  & Training Volume & Recovery Time & Minutes Med Activity & MET Medium & Target Calories & Heart Rate Variability & Remains & Pressure max & Location Change \\ \hline
  
  &   & Daily Movement  & Inactivity Alerts  & Minutes High Activity  & MET High  & Target Miles  & Heart Rate Variability std  & Walk Steps  &   & \\ \hline

\end{tabular}
\caption{List of collected features based on the modality (device) and objectives. MET metabolic-equivalent minutes. For further information about ring and watch features please refer to their online documentation \cite{cite16,cite17,cite18} }
\label{table:1}
\end{table*}
\subsection{Features and Labels}

Considering the fully objective and portable affect assessment in this study, we only used commercial monitoring devices (i.e., Oura ring, Samsung Watch, and Android Phones) and obtained three data modalities using the mentioned devices. We present below an overview of collectable features from these modalities (Table 1 shows the full list of all features). 

\begin{figure}[!th]
    \centering
    \includegraphics[width=\linewidth]{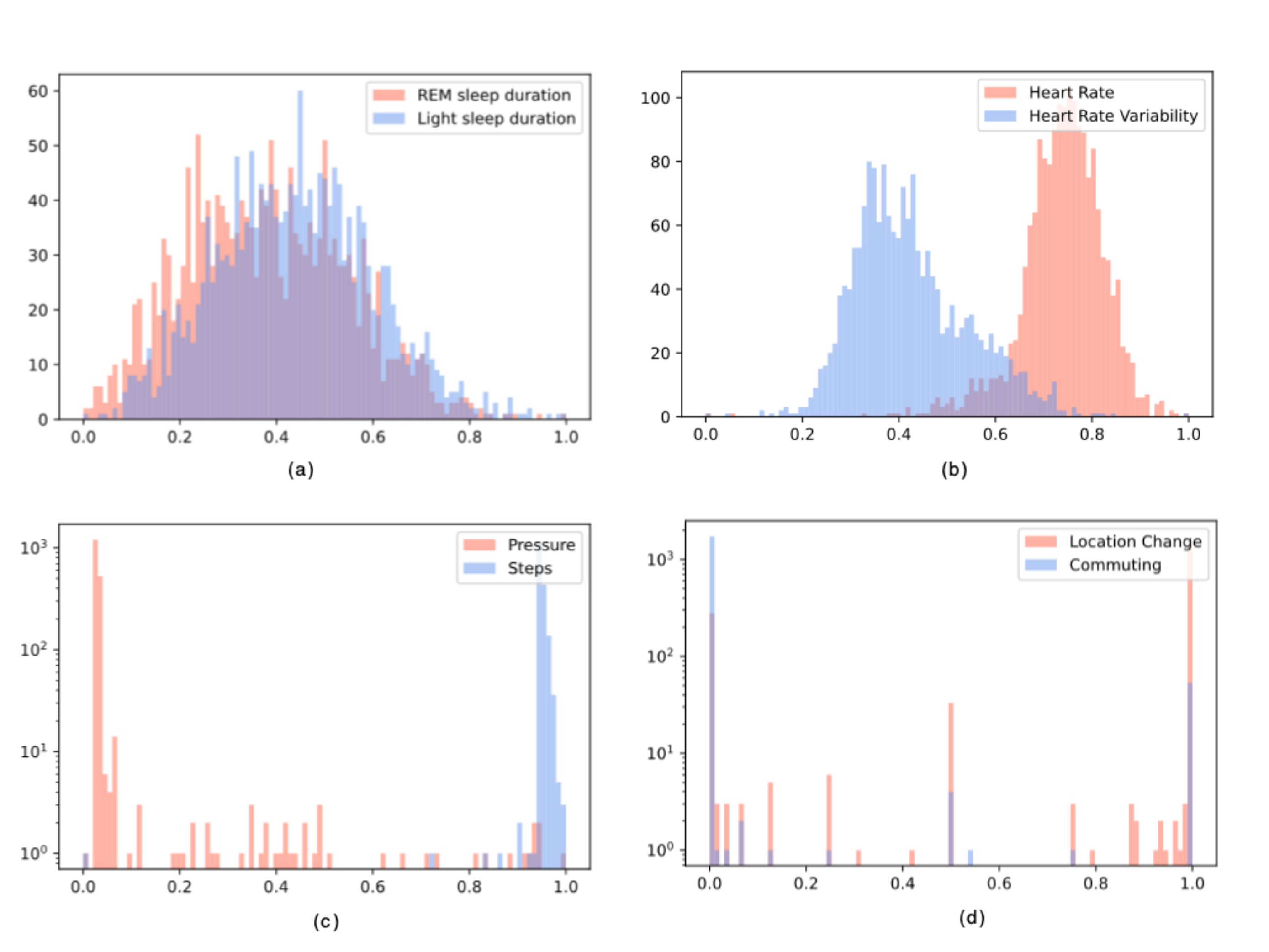}%
    \caption{Distribution of selected features. (a) REM and Light sleep durations, (b) heart rate and heartrate variability collected by ring. (c) atmospheric pressure (altitude) and step counts collected by smartwatch. (d) location change and commuting events detected by Personicle app on the phone.}
    \label{fig:feats_dist}
\end{figure}

\begin{itemize}
  \item \textbf{Smart Ring:}  Oura Ring \cite{cite11} is a smart ring that is capable of extracting meaningful information about sleep (e.g., length of awake, deep and REM sleep stages), physiology (e.g., heart rate, heart rate variability), and activity of users (e.g., daily movement and rest time, etc.). From the viewpoint of  missing data, the convenience of wearing the ring and Oura’s built-in data management makes this modality more continuous and reliable.

  \item \textbf{Smart Watch:} The Samsung smartwatch is another wearable device used for capturing physiological data from participants. Accelerometer (ACC) and photoplethysmography (PPG) are the main measures captured in the smartwatch that are indicators of a user’s physical activities (e.g., steps, movements) and the autonomic nervous system’s activities (e.g., heart rate, heart rate variability). In contrast to smart rings, smart watches are capable of recording the signals with a higher frequency and resolution. However, this results in higher power consumption, the need for more frequent charging, and therefore, greater likelihood of missing data. 
  \item \textbf{Smart Phone:} As mentioned, we also extracted information regarding type of activities and location of the participant through Personicle android monitoring application \cite{cite18} installed on participants’ phone. Using the Personicle app, we collected data on major physical (e.g., in vehicle, still, on bicycle) and behavioral activities (e.g., working, commuting, relaxing) throughout the day. This modality relies on smartphone sensors and applications, and mostly on movement and location to detect these activities; it is possible that due to the lockdown situation and movement limitations of individuals, most of the in-house activities were not detected. This information is available through the Android API which was collected by the phones’ physical sensors and location tracking.
  \item \textbf{Affect (Daily Assessment):} For the daily assessments of affect, participants rated on a scale of 0-100 (0 = “Very Slightly; 100 = “Extremely”) how they felt on a series of 20 different discrete emotion items (e.g., “inspired”, “enthusiastic”, “nervous”, “upset”) over the course of the day. The selected items were adapted from the Positive and Negative Affect Schedule \cite{cite19} , a frequently used scale to assess emotions. Each emotion was examined separately but also as a composite, such that Positive Affect (\textbf{PA}) was calculated as an average of the included 10 PA items (M=45.27, SD=20.22, $\alpha$ = .85) and Negative Affect (\textbf{NA}) was calculated as an average of the included 10 negative affect items (M=21.79, SD=12.28, $\alpha$ = .91) .

\end{itemize}
Finally, using these three modalities, we collected \textbf{52 features}  as shown in Table 1. Some of the features were captured daily whereas some features were captured intensively (e.g. 5 minutes sliding average for ring) over the course of the day. Those captured multiple times a day were weighted-averaged, with respect to the duration.

\subsection{Distributions}
Features representing different biological or behavioral aspects of participants had very different distributions. Figure \ref{fig:feats_dist} (a) and (b) represent (min/max normalized) distributions of sleep and heart related features collected by the Oura ring. Sleep features had a more normal distribution as compared to heart-rate features and in gener`al, values collected by the ring had distributions closer to normal. Figure \ref{fig:feats_dist} (c) depicts atmospheric pressure and number of steps detected by the smartwatch. As we can see, the distribution of the number of steps after min-max normalization is very concentrated (y axis is logarithmic), which could be an effect of the COVID lockdown circumstance. Finally, Figure \ref{fig:feats_dist} (d) shows distributions of selected features from the phone through the Personicle application. Reliance of this modality to GPS and motion sensors made its data largely affected by the pandemic lockdown situation.

\begin{figure}[!th]
    \centering
    \includegraphics[width=\linewidth]{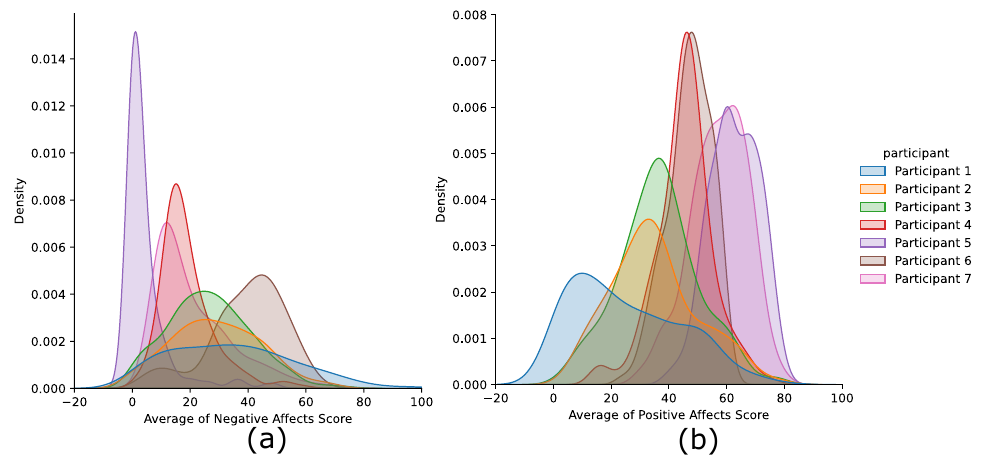}%
    \caption{Smoothed distribution of averaged positive and negative affect of selected seven participants. NAs tend to have a narrower distribution compared to PA.}
    \label{fig:label_dists}
\end{figure}

Figure \ref{fig:label_dists} demonstrates the averaged affect distributions for participants with more than 200 days of valid  affect values. As we can see, PA and NA follow a roughly normal distribution with some possible skews and bumps and diversity in peak values and locations. In general, NA has narrower distribution with higher peaks, this makes NA values labels less discriminative compared to PA. For example, in NA, two days with similar feature values could have small differences but are labeled differently because they are located on different sides of the median (refer to section \ref{section:classification_scheme} for classification scheme).

\subsection{Distributions}

\begin{figure}
    \centering
    \includegraphics[width=\linewidth]{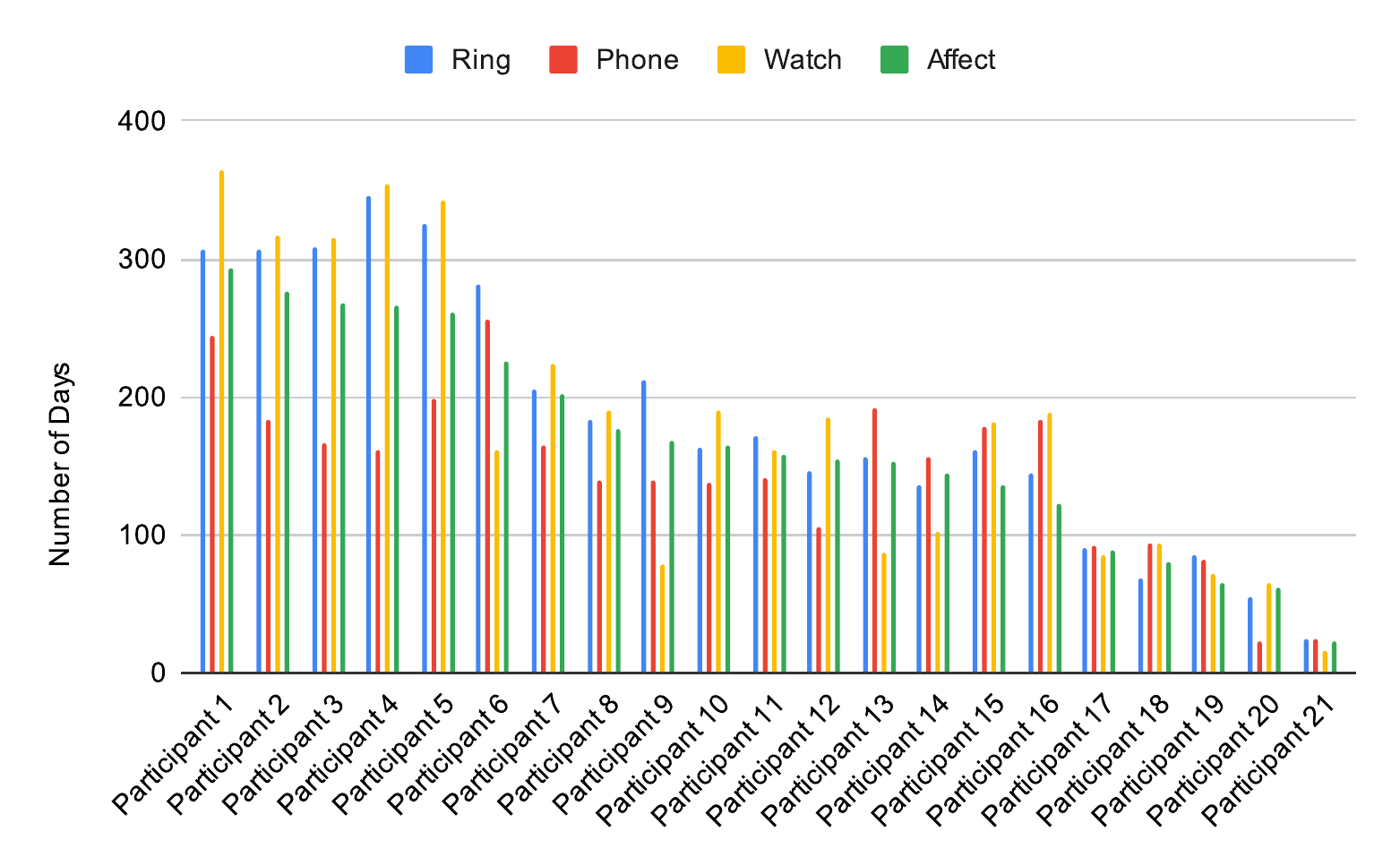}%
    \caption{Participants’ number of days with valid data collected from different devices and EMA sorted by affect values.}
    \label{fig:missing_dist}
\end{figure}

Managing a one-year human subject study in everyday settings as well as  during a pandemic raised several challenges. In particular, the increased volume of missing data poses additional challenges for data modeling. Figure \ref{fig:missing_dist} shows the number of days of data available for each participant and modality. We found that activities detected by phone data (Personicle) was the modality with the highest number of missing values. We speculate that with participants spending most of their time at home (self-quarantining), movements and location detection suffered (e.g., leaving the phone at one’s desk), resulting in poor performance in activity detection. Rings and watches generally had the most available data, confirming the necessity of objective measurements through wearable devices.

For every participant's missing feature value, we averaged values of that feature within two nearest days before and after that date in a 5-day window. No data imputation was applied on the targeted affect values to avoid inaccuracy in the prediction labels. 

\subsection{Predictive Modeling Scheme}
\label{section:classification_scheme}
For prediction, we selected participants with more than 200 days of affect data available. This included 7 of the 20 participants. The classification task is defined as predicting if each affect value is above or below the median of the whole distribution. This is done after removing the middle 20\% of the values in the whole distribution. We used random forests (RF), support vector machine (SVM), multilayer perceptron (MLP) and K-nearest neighbor (KNN) models for training on all features. Hyper parameters of each model were tuned with respect to the accuracy on the 10\% held-out portion of the training (validation set). In general, RF outperformed the other models due to its capability in learning large feature space. Details of performances are available in results and discussion. 

During preprocessing detected activities by Personicle were fed to the model as Boolean variables each representing the weighted average of that feature during the day. Features were z-normalized with respect to the mean and variance of the training set. Random 5 fold cross validation was performed to avoid test set contamination and the accuracies were averaged out.

\section{Results and Discussion}
\begin{figure}
    \centering
    \includegraphics[width=\linewidth]{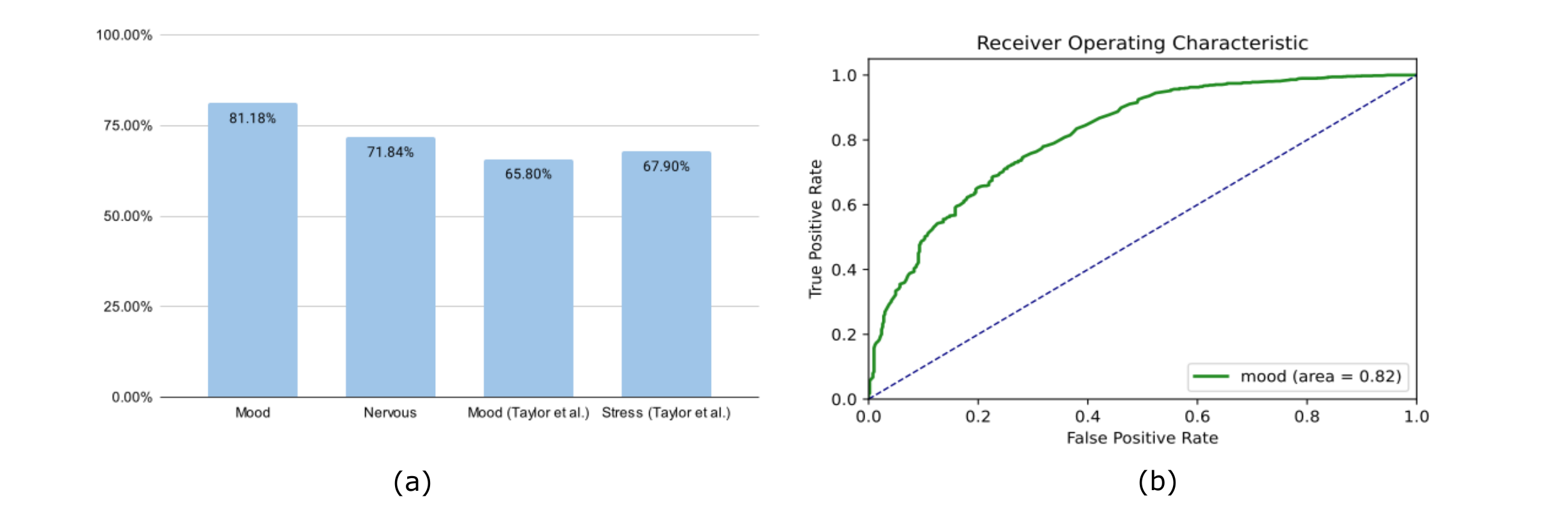}%
    \caption{(a) Accuracy of Random Forest classifier on Nervous affect and compiled mood scores from all affects compared to the models with best performance on SNAPSHOT by Taylor et al. (b) ROC and AUC for mood scores prediction }
    \label{fig:acc_roc}
\end{figure}

\begin{figure}
    \centering
    \includegraphics[width=\linewidth]{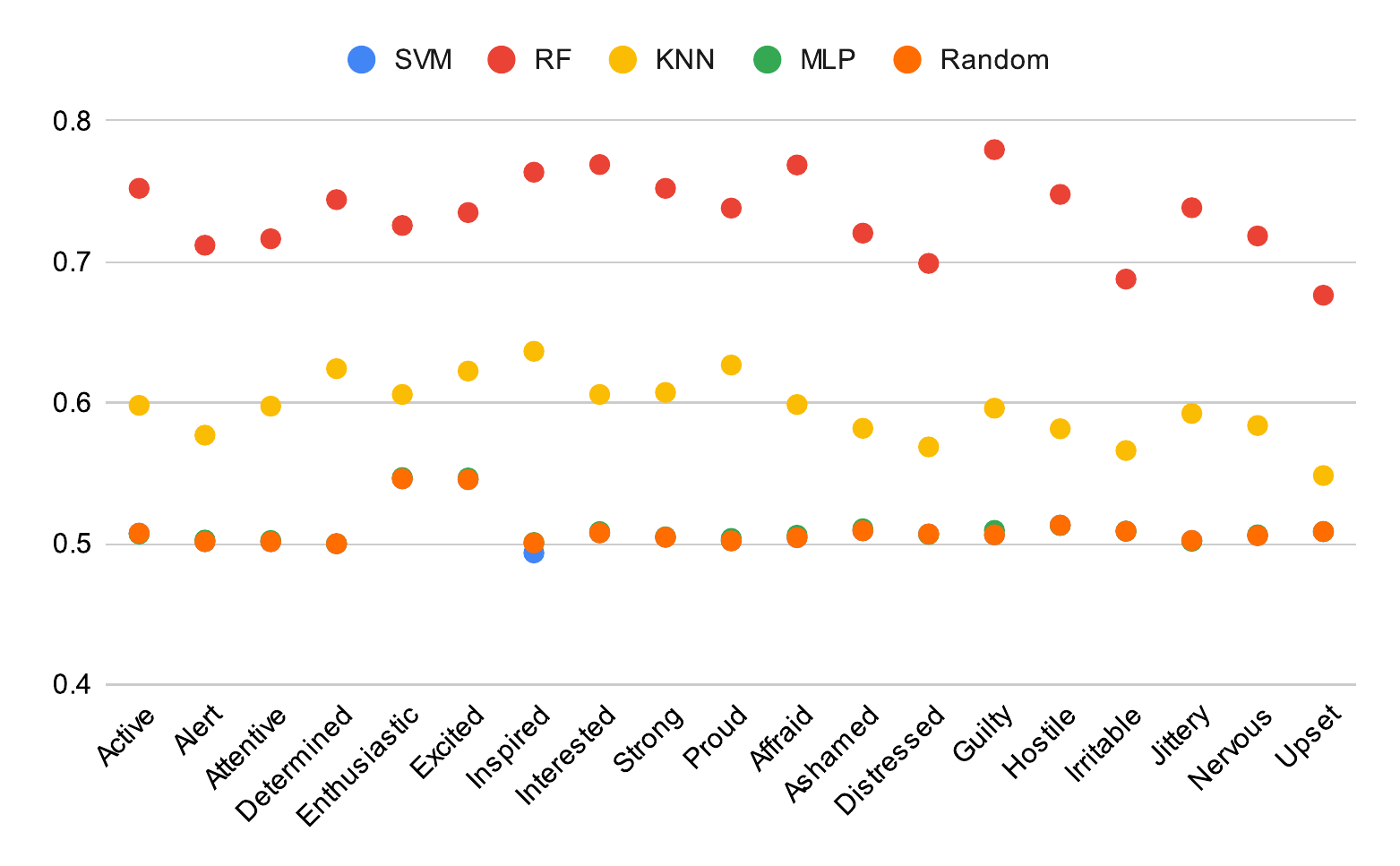}%
    \caption{ Accuracy of RF, SVM, KNN, MLP models and random prediction(majority class) calculated in 5 fold cross validation.
 }
    \label{fig:all_models}
\end{figure}

We compared our results to two of the state-of-the-art studies conducted on the SNAPSHOT dataset,Taylor et al. and Spathis et. al \cite{cite12}. In order to be able to compare our models with other related works in which an aggregate single binary value is used to represent  mood (i.e., sad/happy), we calculated the deviation of the PA and NA values from their median of the entire distribution for each participant, and selected the one with a larger deviation as the dominant mood and assigned it as the final (sad/happy) mood label. Our general model for predicting mood and nervousness outperformed the state-of-the-art in predicting mood and stress. The Random Forest model was selected for comparison here. ( detailed performance values of the other models are shown in Figure \ref{fig:all_models}). Figure \ref{fig:acc_roc}-(a) compares prediction accuracy of mood and nervousness affect  to Taylor et al \cite{cite6} showing about 16\% and 4\% improvement in accuracy for mood and stress (i.e., nervousness) prediction, respectively. Figure \ref{fig:acc_roc}-(b) shows the receiver operating characteristic (ROC) of the mood prediction model yielding 0.82 area under curve (AUC) that has 8.1\% improvement compared to the best values reported by Spathis et. al  for mood prediction. 

\begin{figure}
    \centering
    \includegraphics[width=\linewidth]{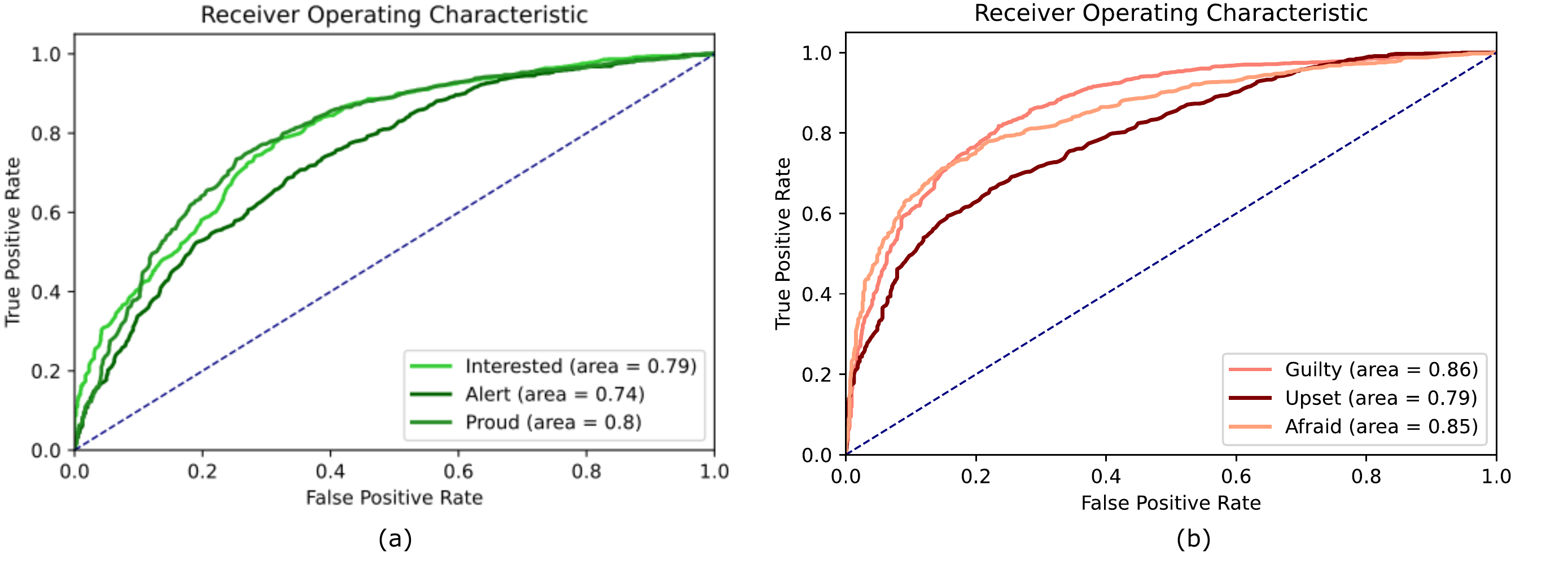}%
    \caption{(a) Accuracy of RF classifier on Nervous affect and compiled mood scores from all affects compared to the models with best performance on SNAPSHOT by Taylor et al. (b) ROC and AUC for mood scores prediction }
    \label{fig:roc_pos_neg}
\end{figure}

Figure \ref{fig:roc_pos_neg} (a) and (b) also show the highest, lowest, and median of PA and NA with respect to their AUCs. NA models tends to have a higher ROC but in general lower accuracy (Please see Figure \ref{fig:all_models} for detailed results). The main reason is the comparatively narrower distribution of the NA (Figure \ref{fig:label_dists}) that makes the close values get different labels. This indicates that despite better probabilities being generated for NA, the chance of producing the wrong label is higher since labeling thresholds are narrower.

\begin{figure}
    \centering
    \includegraphics[width=\linewidth]{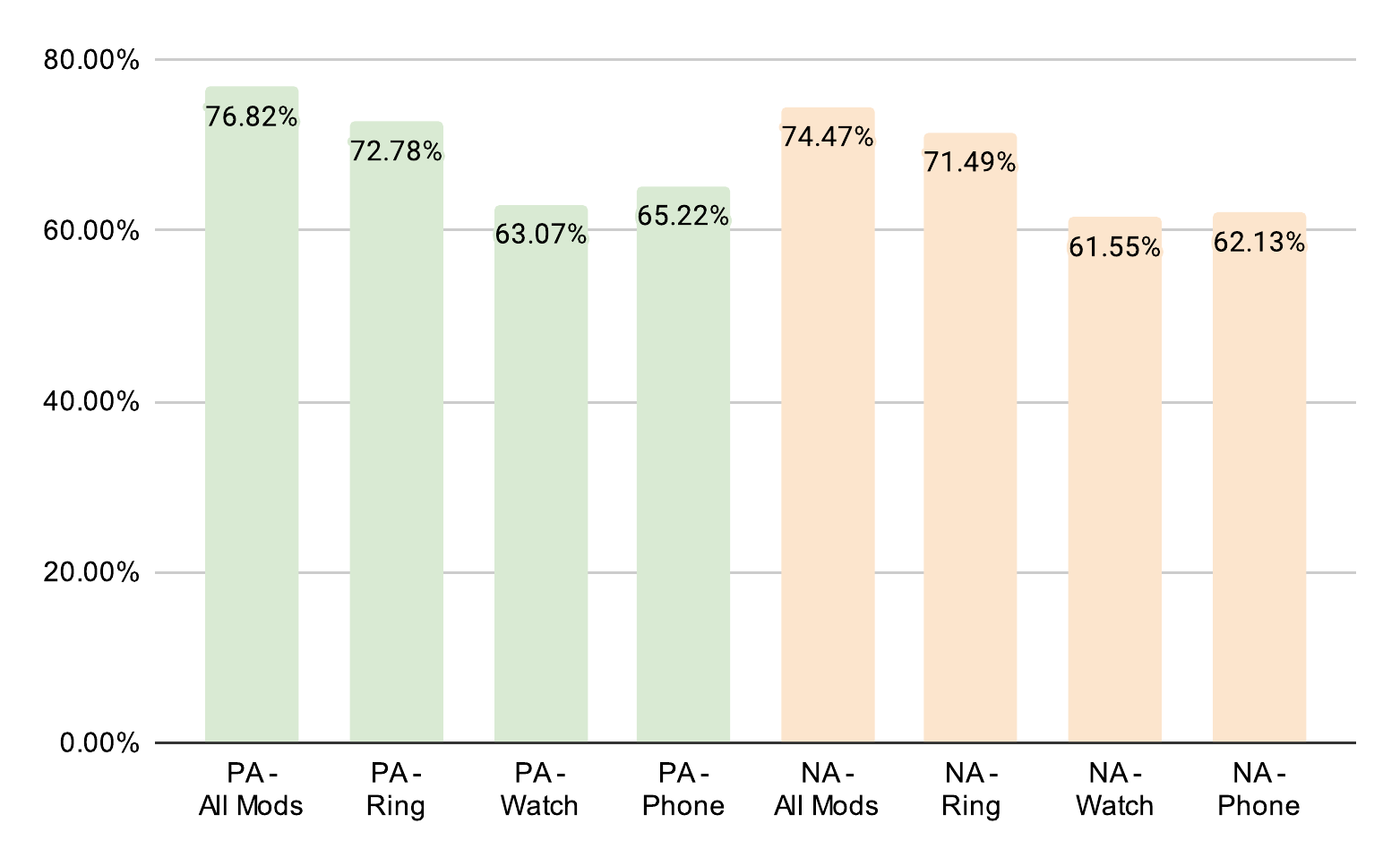}%
    \caption{ Performance of RF trained on modalities separately and all together for PA and NA }
    \label{fig:acc_mod}
\end{figure}

One of the objectives of this study was to explore capabilities of leveraging multimodal data for modeling affect. In general, multimodal machine learning could be used to either increase robustness (i.e., providing redundant information) or improve prediction performance (i.e., obtaining additional information from different aspects of the event). In this study, we focused on the later purpose of multimodal machine learning and therefore, assessment devices were selected in order to cover different aspects of an individual's life from physiological (ring and watch) to behavioral (phone). To test the power of multimodal assessment, we trained models modalities separately and all together. Figure \ref{fig:acc_mod} compares accuracies of models trained with data from different devices separately to ones trained with data from all devices for NA and PA. As shown, leveraging all of the modalities improve accuracy by up to 21.8\%.

\begin{figure}
    \centering
    \includegraphics[width=\linewidth]{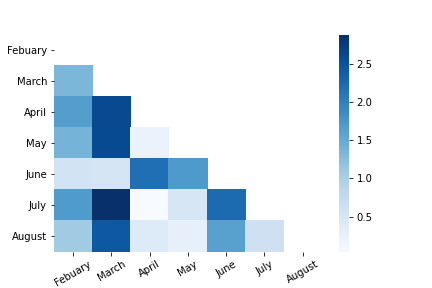}%
    \caption{ Performance of RF trained on modalities separately and all together for PA and NA }
    \label{fig:month_heatmap}
\end{figure}

Another unique aspect of this study was its timing--the study began in early 2020 and covered the period before and during the stay-at-home order in the state of California. The stay-at-home order in most of the states happened around March and affected almost everyone’s lives, including college students, in many different ways.  In order to investigate the effects of this event, we used our mood prediction model to predict participants’ mood during the last week of every month. Then we computed t-value as a metric of difference of predicted moods between months before and after lockdown order. Figure \ref{fig:month_heatmap} shows absolute t-values of predicted mood for participants across these 1-month intervals. As we observe, March has the highest t-value (difference from other months) that coincides with the very early stages of lockdown.

\begin{figure}
    \centering
    \includegraphics[width=\linewidth]{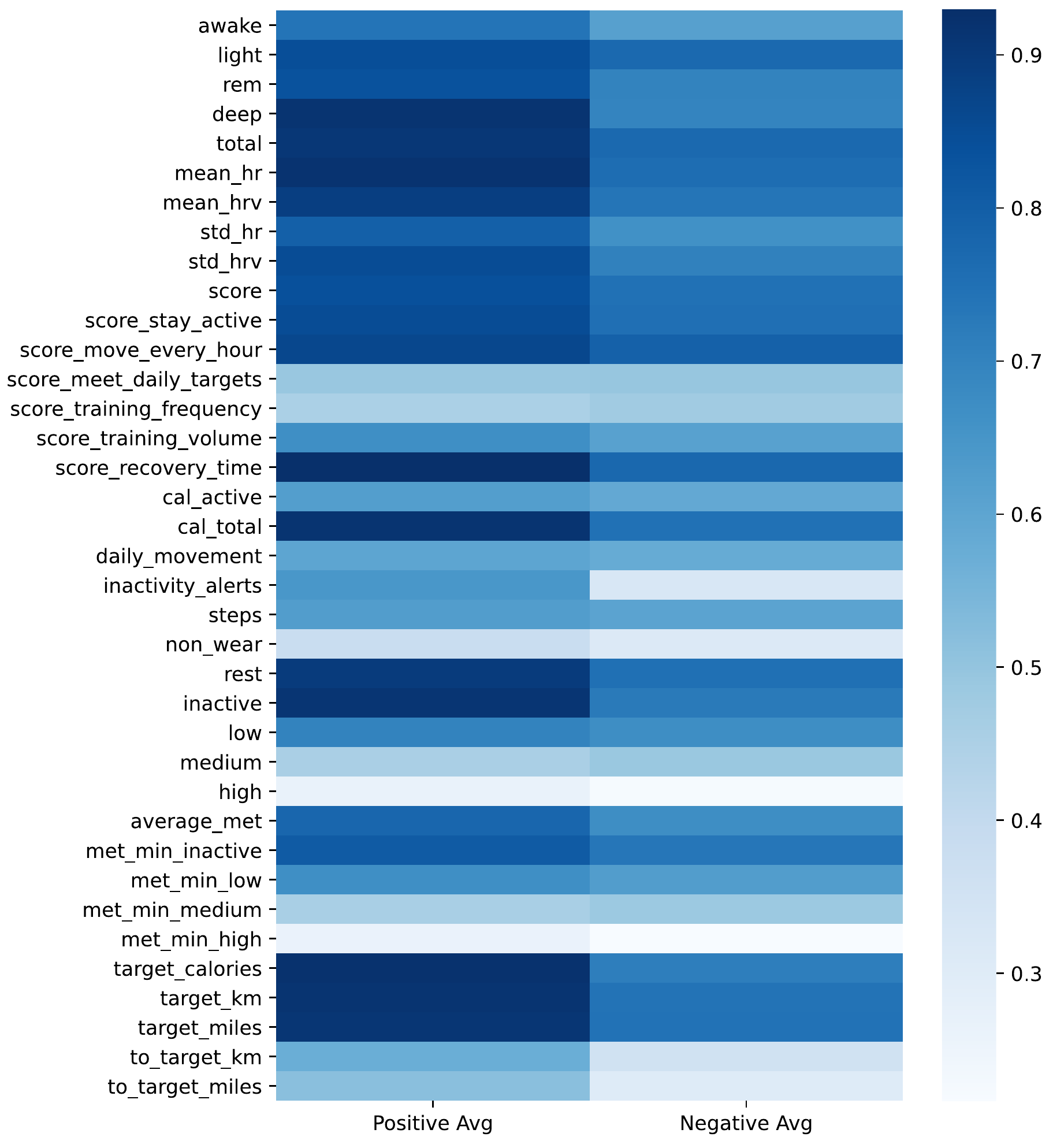}%
    \caption{ Correlation of features collected by objective assessment with two target labels}
    \label{fig:corr_heatmap}
\end{figure}

To gain a deeper understanding about different modalities, we conducted correlational analyses to explore associations of physiological and personal chronicle features with PA and NA. Figure \ref{fig:corr_heatmap} (a) shows correlation of ring features to PA and NA. Duration of light and deep sleep along with sleep score and inactivity level were the main indicators of PA, while NA was mainly correlated with activity-related features. The findings of this study resonate with prior work suggesting that sleep may predict PA. PA may be one potential pathway through which sleep has implications for mental and physical health outcomes. Specifically, individuals with sleep disturbances report more blunted positive affect associated with rewarding activities such as decreased energy and happiness compared to individuals with extended sleep \cite{cite20}. 

We also conducted more detailed experiment on different models and trained different machine learning models on all the specific reported affect features separately. Figure \ref{fig:all_models} demonstrates the results for all the experiments. Random forests (RF) outperformed all the other models in terms of accuracy.

\section{Conclusion and Future Work}

This study was designed to investigate the viability of predicting an individual's mood and affects only by leveraging objective measurements without a need to intervene users for collecting feedback. We conducted this study over the course of one year which is a wider time span compared to similar works.  The chance of observing young adults during the lockdown situation of COVID-19 pandemic gives a unique dimension to this study. Different watch, phone and web applications were developed to objectively assess participants using commercially available devices. Results showed that we can predict next-day mood with accuracy of ~81\% using generic machine learning algorithms. These models outperformed their alternatives in similar studies even without an individual’s historical information for bringing personalization or temporal context. Results show that  sleep and heart rate variability features collected by the smart ring are among the best predictors of mood and positive affect which suggests further investigation. 

During the study COVID-19 lockdown circumstances imposed some challenges on data collection due to the limited access to the participants and their normal movement and commute patterns. These limitations made missing data a challenging part for building predictive models. We observed that generic machine learning models on our dataset resulted in comparable performance to counterpart studies even using personalization and temporal context. 

In our future studies we plan to investigate novel methods that incorporate patient-specific information into modeling and explore high-resolution signals (raw photoplethysmography, accelerometer) collected by wearables. Additionally, monitoring and predicting affect, and eventually developing and refining interventions to regulate and manage affect using technology is an important future avenue. Specifically, the findings of this study are important in showing that the advancements in objective monitoring and predicting affect using features such as sleep may provide important opportunities in a less burdensome approach of monitoring and predicting important aspects of individuals’ health. Future research may benefit from designing and refining interventions among emerging adults using smartphones. For instance, monitoring and prediction of affect could be used to identify the moments to use interventions in hopes of personalization of interventions targeting affect and other aspects of mental health.

\bibliographystyle{unsrt}
\bibliography{citations}

\begin{thebibliography}{10}

\bibitem{cite1}
Mental illness.
\newblock \url{https://www.nimh.nih.gov/health/statistics/mental-illness}.

\bibitem{cite2}
Lisa~Feldman Barrett and Eliza Bliss-Moreau.
\newblock Affect as a psychological primitive.
\newblock {\em Advances in experimental social psychology}, 41:167--218, 2009.

\bibitem{cite3}
James~J Gross and Hooria Jazaieri.
\newblock Emotion, emotion regulation, and psychopathology: An affective
  science perspective.
\newblock {\em Clinical psychological science}, 2(4):387--401, 2014.

\bibitem{cite4}
James~W Pennebaker.
\newblock Writing about emotional experiences as a therapeutic process.
\newblock {\em Psychological science}, 8(3):162--166, 1997.

\bibitem{cite5}
Mor Nahum, Thomas~M Van~Vleet, Vikaas~S Sohal, Julie~J Mirzabekov, Vikram~R
  Rao, Deanna~L Wallace, Morgan~B Lee, Heather Dawes, Alit Stark-Inbar,
  Joshua~Thomas Jordan, et~al.
\newblock Immediate mood scaler: tracking symptoms of depression and anxiety
  using a novel mobile mood scale.
\newblock {\em JMIR mHealth and uHealth}, 5(4):e6544, 2017.

\bibitem{cite6}
Sara Taylor, Natasha Jaques, Ehimwenma Nosakhare, Akane Sano, and Rosalind
  Picard.
\newblock Personalized multitask learning for predicting tomorrow's mood,
  stress, and health.
\newblock {\em IEEE Transactions on Affective Computing}, 11(2):200--213, 2017.

\bibitem{cite8}
U.S.~Census Bureau.
\newblock Initial impact of covid-19 on u.s. economy more widespread than on
  mortality.
\newblock
  \url{https://www.census.gov/library/stories/2021/03/initial-impact-covid-19-on-united-states-economy-more-widespread-than-on-mortality.html},
  Mar 2021.

\bibitem{cite14}
Amir~M Rahmani, Jocelyn Lai, Salar Jafarlou, Asal Yunusova, Alex Rivera, Sina
  Labbaf, Sirui Hu, Arman Anzanpour, Nikil Dutt, Ramesh Jain, et~al.
\newblock Personal mental health navigator: Harnessing the power of data,
  personal models, and health cybernetics to promote psychological well-being.
\newblock {\em arXiv preprint arXiv:2012.09131}, 2020.

\bibitem{cite9}
Jocelyn Lai, Amir Rahmani, Asal Yunusova, Alexander~P Rivera, Sina Labbaf,
  Sirui Hu, Nikil Dutt, Ramesh Jain, Jessica~L Borelli, et~al.
\newblock Using multimodal assessments to capture personalized contexts of
  college student well-being in 2020: Case study.
\newblock {\em JMIR formative research}, 5(5):e26186, 2021.

\bibitem{cite10}
Asal Yunusova, Jocelyn Lai, Alexander~P Rivera, Sirui Hu, Sina Labbaf, Amir~M
  Rahmani, Nikil Dutt, Ramesh~C Jain, and Jessica~L Borelli.
\newblock Assessing the mental health of emerging adults through a mental
  health app: Protocol for a prospective pilot study.
\newblock {\em JMIR Research Protocols}, 10(3):e25775, 2021.

\bibitem{cite13}
Akane Sano.
\newblock {\em Measuring college students sleep, stress and mental health with
  wearable sensors and mobile phones}.
\newblock PhD thesis, PhD thesis, MIT, 2015.

\bibitem{cite7}
Timothy~J Trull and Ulrich Ebner-Priemer.
\newblock Ambulatory assessment.
\newblock {\em Annual review of clinical psychology}, 9:151--176, 2013.

\bibitem{cite20}
Zahra Mousavi, Jocelyn Lai, Asal Yunusova, Alexander Rivera, Sirui Hu, Sina
  Labbaf, Salar Jafarlou, Nikil Dutt, Ramesh Jain, Amir Rahmani, et~al.
\newblock 193 sleep variability and affect dynamics among college students
  during covid-19 pandemic.
\newblock {\em Sleep}, 44(Supplement\_2):A78--A78, 2021.

\bibitem{cite15}
Arman Anzanpour.
\newblock Health scitech group: Zotcare.
\newblock \url{http://healthscitech.nursing.uci.edu/projects/zotcare/}.

\bibitem{cite16}
Api documentation.
\newblock \url{https://cloud.ouraring.com/docs/}.

\bibitem{cite17}
Galaxy watch - build.
\newblock \url{https://developer.samsung.com/galaxy-watch-develop}.

\bibitem{cite18}
Laleh Jalali, Da~Huo, Hyungik Oh, Mengfan Tang, Siripen Pongpaichet, and Ramesh
  Jain.
\newblock Personicle: personal chronicle of life events.
\newblock In {\em Workshop on Personal Data Analytics in the Internet of Things
  (PDA@ IOT) at the 40th International Conference on Very Large Databases
  (VLDB), Hangzhou, China}, 2014.

\bibitem{cite11}
Oura Ring.
\newblock Accurate health information accessible to everyone.
\newblock {\em Oura. URL: https://ouraring. com/[accessed 2021-02-02]}.

\bibitem{cite19}
David Watson, Lee~Anna Clark, and Auke Tellegen.
\newblock Development and validation of brief measures of positive and negative
  affect: the panas scales.
\newblock {\em Journal of personality and social psychology}, 54(6):1063, 1988.

\bibitem{cite12}
Dimitris Spathis, Sandra Servia-Rodriguez, Katayoun Farrahi, Cecilia Mascolo,
  and Jason Rentfrow.
\newblock Passive mobile sensing and psychological traits for large scale mood
  prediction.
\newblock In {\em Proceedings of the 13th EAI International Conference on
  Pervasive Computing Technologies for Healthcare}, pages 272--281, 2019.

\end{thebibliography}

\end{document}